\expandafter\ifx\csname phyzzx\endcsname\relax\else
 \errhelp{Hit <CR> and go ahead.}
 \errmessage{PHYZZX macros are already loaded or input. }
 \endinput \fi
\catcode`\@=11 
%
%
%
\font\seventeenrm=cmr17
\font\fourteenrm=cmr12 scaled\magstep1
\font\twelverm=cmr12
\font\ninerm=cmr9            \font\sixrm=cmr6
%
\font\fourteenbf=cmbx10 scaled\magstep2
\font\twelvebf=cmbx12
\font\ninebf=cmbx9            \font\sixbf=cmbx6
%
\font\fourteeni=cmmi10 scaled\magstep2      \skewchar\fourteeni='177
\font\twelvei=cmmi12			        \skewchar\twelvei='177
\font\ninei=cmmi9                           \skewchar\ninei='177
\font\sixi=cmmi6                            \skewchar\sixi='177
%
\font\fourteensy=cmsy10 scaled\magstep2     \skewchar\fourteensy='60
\font\twelvesy=cmsy10 scaled\magstep1	    \skewchar\twelvesy='60
\font\ninesy=cmsy9                          \skewchar\ninesy='60
\font\sixsy=cmsy6                           \skewchar\sixsy='60
%
\font\fourteenex=cmex10 scaled\magstep2
\font\twelveex=cmex10 scaled\magstep1
%
\font\fourteensl=cmsl12 scaled\magstep1
\font\twelvesl=cmsl12
\font\ninesl=cmsl9
%
\font\fourteenit=cmti12 scaled\magstep1
\font\twelveit=cmti12
\font\nineit=cmti9
\font\fourteentt=cmtt10 scaled\magstep2
\font\twelvett=cmtt12
\font\fourteencp=cmcsc10 scaled\magstep2
\font\twelvecp=cmcsc10 scaled\magstep1
\font\tencp=cmcsc10
\newfam\cpfam
\newdimen\b@gheight		\b@gheight=12pt
\newcount\f@ntkey		\f@ntkey=0
\def\f@m{\afterassignment\samef@nt\f@ntkey=}
\def\samef@nt{\fam=\f@ntkey \the\textfont\f@ntkey\relax}
\def\rm{\f@m0 }
\def\mit{\f@m1 }         
\def\cal{\f@m2 }
\def\it{\f@m\itfam}
\def\sl{\f@m\slfam}
\def\bf{\f@m\bffam}
\def\tt{\f@m\ttfam}
\def\caps{\f@m\cpfam}
\def\fourteenpoint{\relax
    \textfont0=\fourteenrm          \scriptfont0=\tenrm
      \scriptscriptfont0=\sevenrm
    \textfont1=\fourteeni           \scriptfont1=\teni
      \scriptscriptfont1=\seveni
    \textfont2=\fourteensy          \scriptfont2=\tensy
      \scriptscriptfont2=\sevensy
    \textfont3=\fourteenex          \scriptfont3=\twelveex
      \scriptscriptfont3=\tenex
    \textfont\itfam=\fourteenit     \scriptfont\itfam=\tenit
    \textfont\slfam=\fourteensl     \scriptfont\slfam=\tensl
    \textfont\bffam=\fourteenbf     \scriptfont\bffam=\tenbf
      \scriptscriptfont\bffam=\sevenbf
    \textfont\ttfam=\fourteentt
    \textfont\cpfam=\fourteencp
    \samef@nt
    \b@gheight=14pt
    \setbox\strutbox=\hbox{\vrule height 0.85\b@gheight
				depth 0.35\b@gheight width\z@ }}
\def\twelvepoint{\relax
    \textfont0=\twelverm          \scriptfont0=\ninerm
      \scriptscriptfont0=\sixrm
    \textfont1=\twelvei           \scriptfont1=\ninei
      \scriptscriptfont1=\sixi
    \textfont2=\twelvesy           \scriptfont2=\ninesy
      \scriptscriptfont2=\sixsy
    \textfont3=\twelveex          \scriptfont3=\tenex
      \scriptscriptfont3=\tenex
    \textfont\itfam=\twelveit     \scriptfont\itfam=\nineit
    \textfont\slfam=\twelvesl     \scriptfont\slfam=\ninesl
    \textfont\bffam=\twelvebf     \scriptfont\bffam=\ninebf
      \scriptscriptfont\bffam=\sixbf
    \textfont\ttfam=\twelvett
    \textfont\cpfam=\twelvecp
    \samef@nt
    \b@gheight=12pt
    \setbox\strutbox=\hbox{\vrule height 0.85\b@gheight
				depth 0.35\b@gheight width\z@ }}
\def\tenpoint{\relax
    \textfont0=\tenrm          \scriptfont0=\sevenrm
      \scriptscriptfont0=\fiverm
    \textfont1=\teni           \scriptfont1=\seveni
      \scriptscriptfont1=\fivei
    \textfont2=\tensy          \scriptfont2=\sevensy
      \scriptscriptfont2=\fivesy
    \textfont3=\tenex          \scriptfont3=\tenex
      \scriptscriptfont3=\tenex
    \textfont\itfam=\tenit     \scriptfont\itfam=\seveni
    \textfont\slfam=\tensl     \scriptfont\slfam=\sevenrm
    \textfont\bffam=\tenbf     \scriptfont\bffam=\sevenbf
      \scriptscriptfont\bffam=\fivebf
    \textfont\ttfam=\tentt
    \textfont\cpfam=\tencp
    \samef@nt
    \b@gheight=10pt
    \setbox\strutbox=\hbox{\vrule height 0.85\b@gheight
				depth 0.35\b@gheight width\z@ }}
%
%
%
\normalbaselineskip = 20pt plus 0.2pt minus 0.1pt
\normallineskip = 1.5pt plus 0.1pt minus 0.1pt
\normallineskiplimit = 1.5pt
\newskip\normaldisplayskip
\normaldisplayskip = 20pt plus 5pt minus 10pt
\newskip\normaldispshortskip
\normaldispshortskip = 6pt plus 5pt
\newskip\normalparskip
\normalparskip = 6pt plus 2pt minus 1pt
\newskip\skipregister
\skipregister = 5pt plus 2pt minus 1.5pt
\newif\ifsingl@    \newif\ifdoubl@
\newif\iftwelv@    \twelv@true
\def\singlespace{\singl@true\doubl@false\spaces@t}
\def\doublespace{\singl@false\doubl@true\spaces@t}
\def\normalspace{\singl@false\doubl@false\spaces@t}
\def\Tenpoint{\tenpoint\twelv@false\spaces@t}
\def\Twelvepoint{\twelvepoint\twelv@true\spaces@t}
\def\spaces@t{\relax
      \iftwelv@ \ifsingl@\subspaces@t3:4;\else\subspaces@t1:1;\fi
       \else \ifsingl@\subspaces@t3:5;\else\subspaces@t4:5;\fi \fi
      \ifdoubl@ \multiply\baselineskip by 5
         \divide\baselineskip by 4 \fi }
\def\subspaces@t#1:#2;{
      \baselineskip = \normalbaselineskip
      \multiply\baselineskip by #1 \divide\baselineskip by #2
      \lineskip = \normallineskip
      \multiply\lineskip by #1 \divide\lineskip by #2
      \lineskiplimit = \normallineskiplimit
      \multiply\lineskiplimit by #1 \divide\lineskiplimit by #2
      \parskip = \normalparskip
      \multiply\parskip by #1 \divide\parskip by #2
      \abovedisplayskip = \normaldisplayskip
      \multiply\abovedisplayskip by #1 \divide\abovedisplayskip by #2
      \belowdisplayskip = \abovedisplayskip
      \abovedisplayshortskip = \normaldispshortskip
      \multiply\abovedisplayshortskip by #1
        \divide\abovedisplayshortskip by #2
      \belowdisplayshortskip = \abovedisplayshortskip
      \advance\belowdisplayshortskip by \belowdisplayskip
      \divide\belowdisplayshortskip by 2
      \smallskipamount = \skipregister
      \multiply\smallskipamount by #1 \divide\smallskipamount by #2
      \medskipamount = \smallskipamount \multiply\medskipamount by 2
      \bigskipamount = \smallskipamount \multiply\bigskipamount by 4 }
\def\normalbaselines{ \baselineskip=\normalbaselineskip
   \lineskip=\normallineskip \lineskiplimit=\normallineskip
   \iftwelv@\else \multiply\baselineskip by 4 \divide\baselineskip by 5
     \multiply\lineskiplimit by 4 \divide\lineskiplimit by 5
     \multiply\lineskip by 4 \divide\lineskip by 5 \fi }
\Twelvepoint  
\interlinepenalty=50
\interfootnotelinepenalty=5000
\predisplaypenalty=9000
\postdisplaypenalty=500
\hfuzz=1pt
\vfuzz=0.2pt
\voffset=0pt
\dimen\footins=8 truein
%
%
%
\def\pagecontents{
   \ifvoid\topins\else\unvbox\topins\vskip\skip\topins\fi
   \dimen@ = \dp255 \unvbox255
   \ifvoid\footins\else\vskip\skip\footins\footrule\unvbox\footins\fi
   \ifr@ggedbottom \kern-\dimen@ \vfil \fi }
\def\makeheadline{\vbox to 0pt{ \skip@=\topskip
      \advance\skip@ by -12pt \advance\skip@ by -2\normalbaselineskip
      \vskip\skip@ \line{\vbox to 12pt{}\the\headline} \vss
      }\nointerlineskip}
\def\makefootline{\baselineskip = 1.5\normalbaselineskip
                 \line{\the\footline}}
\newif\iffrontpage
\newif\ifletterstyle
\newif\ifp@genum
\def\nopagenumbers{\p@genumfalse}
\def\pagenumbers{\p@genumtrue}
\pagenumbers
\newtoks\paperheadline
\newtoks\letterheadline
\newtoks\paperfootline
\newtoks\letterfootline
\newtoks\letterinfo
\newtoks\Letterinfo
\newtoks\date
\footline={\ifletterstyle\the\letterfootline\else\the\paperfootline\fi}
\paperfootline={\hss\iffrontpage\else\ifp@genum\tenrm\folio\hss\fi\fi}
\letterfootline={\iffrontpage\LETTERFOOT\else\hfil\fi}
\Letterinfo={\hfil}
\letterinfo={\hfil}
\def\LETTERFOOT{\hfil} 
%
\def\LETTERHEAD{\vtop{\baselineskip=20pt\hbox to
\hsize{\hfil\seventeenrm\strut
CALIFORNIA INSTITUTE OF TECHNOLOGY \hfil}
\hbox to \hsize{\hfil\ninerm\strut
CHARLES C. LAURITSEN LABORATORY OF HIGH ENERGY PHYSICS \hfil}
\hbox to \hsize{\hfil\ninerm\strut
PASADENA, CALIFORNIA 91125 \hfil}}}
\headline={\ifletterstyle\the\letterheadline\else\the\paperheadline\fi}
\paperheadline={\hfil}
\letterheadline{\iffrontpage \LETTERHEAD\else
    \rm \ifp@genum \hfil \folio\hfil\fi\fi}
\def\monthname{\relax\ifcase\month 0/\or January\or February\or
   March\or April\or May\or June\or July\or August\or September\or
   October\or November\or December\else\number\month/\fi}
\def\today{\monthname\ \number\day, \number\year}
\date={\today}
\countdef\pageno=1      \countdef\pagen@=0
\countdef\pagenumber=1  \pagenumber=1
\def\advancepageno{\global\advance\pagen@ by 1
   \ifnum\pagenumber<0 \global\advance\pagenumber by -1
    \else\global\advance\pagenumber by 1 \fi \global\frontpagefalse }
\def\folio{\ifnum\pagenumber<0 \romannumeral-\pagenumber
           \else \number\pagenumber \fi }
\def\footrule{\dimen@=\prevdepth\nointerlineskip
   \vbox to 0pt{\vskip -0.25\baselineskip \hrule width 0.35\hsize \vss}
   \prevdepth=\dimen@ }
\newtoks\foottokens
\foottokens={}
\newdimen\footindent
\footindent=24pt
\def\vfootnote#1{\insert\footins\bgroup
   \interlinepenalty=\interfootnotelinepenalty \floatingpenalty=20000
   \singl@true\doubl@false\Tenpoint
   \splittopskip=\ht\strutbox \boxmaxdepth=\dp\strutbox
   \leftskip=\footindent \rightskip=\z@skip
   \parindent=0.5\footindent \parfillskip=0pt plus 1fil
   \spaceskip=\z@skip \xspaceskip=\z@skip
   \the\foottokens
   \Textindent{$ #1 $}\footstrut\futurelet\next\fo@t}
\def\Textindent#1{\noindent\llap{#1\enspace}\ignorespaces}
\def\footnote#1{\attach{#1}\vfootnote{#1}}

\let\footsymbol=\star
\newcount\lastf@@t           \lastf@@t=-1
\newcount\footsymbolcount    \footsymbolcount=0
\newif\ifPhysRev
\def\bumpfootsymbolcount{\relax
   \iffrontpage \bumpfootsymbolNP \else \advance\lastf@@t by 1
     \ifPhysRev \bumpfootsymbolPR \else \bumpfootsymbolNP \fi \fi
   \global\lastf@@t=\pagen@ }
\def\bumpfootsymbolNP{\ifnum\footsymbolcount <0 \global\footsymbolcount =0 \fi
    \ifnum\lastf@@t<\pagen@ \global\footsymbolcount=0
     \else \global\advance\footsymbolcount by 1 \fi }
\def\bumpfootsymbolPR{\ifnum\footsymbolcount >0 \global\footsymbolcount =0 \fi
      \global\advance\footsymbolcount by -1 }
\def\fd@f#1 {\xdef\footsymbol{\mathchar"#1 }}
\def\generatefootsymbol{\ifcase\footsymbolcount \fd@f 13F \or \fd@f 279
	\or \fd@f 27A \or \fd@f 278 \or \fd@f 27B \else
	\ifnum\footsymbolcount <0 \fd@f{023 \number-\footsymbolcount }
	 \else \fd@f 203 {\loop \ifnum\footsymbolcount >5
		\fd@f{203 \footsymbol } \advance\footsymbolcount by -1
		\repeat }\fi \fi }

\def\nonfrenchspacing{\sfcode`\.=3001 \sfcode`\!=3000 \sfcode`\?=3000
	\sfcode`\:=2000 \sfcode`\;=1500 \sfcode`\,=1251 }
\nonfrenchspacing
\newdimen\d@twidth
{\setbox0=\hbox{s.} \global\d@twidth=\wd0 \setbox0=\hbox{s}
	\global\advance\d@twidth by -\wd0 }
\def\removehglue{\loop \unskip \ifdim\lastskip >\z@ \repeat }
\def\roll@ver#1{\removehglue \nobreak \count255 =\spacefactor \dimen@=\z@
	\ifnum\count255 =3001 \dimen@=\d@twidth \fi
	\ifnum\count255 =1251 \dimen@=\d@twidth \fi
    \iftwelv@ \kern-\dimen@ \else \kern-0.83\dimen@ \fi
   #1\spacefactor=\count255 }
\def\step@ver#1{\relax \ifmmode #1\else \ifhmode
	\roll@ver{${}#1$}\else {\setbox0=\hbox{${}#1$}}\fi\fi }
\def\attach#1{\step@ver{\strut^{\mkern 2mu #1} }}
%
%
%
\newcount\chapternumber      \chapternumber=0
\newcount\sectionnumber      \sectionnumber=0
\newcount\equanumber         \equanumber=0
\let\chapterlabel=\relax
\let\sectionlabel=\relax
\newtoks\chapterstyle        \chapterstyle={\Number}
\newtoks\sectionstyle        \sectionstyle={\chapterlabel\Number}
\newskip\chapterskip         \chapterskip=\bigskipamount
\newskip\sectionskip         \sectionskip=\medskipamount
\newskip\headskip            \headskip=8pt plus 3pt minus 3pt
\newdimen\chapterminspace    \chapterminspace=15pc
\newdimen\sectionminspace    \sectionminspace=10pc
\newdimen\referenceminspace  \referenceminspace=25pc
\def\chapterreset{\global\advance\chapternumber by 1
   \ifnum\equanumber<0 \else\global\equanumber=0\fi
   \sectionnumber=0 \makechapterlabel}
\def\makechapterlabel{\let\sectionlabel=\relax
   \xdef\chapterlabel{\the\chapterstyle{\the\chapternumber}.}}
\def\alphabetic#1{\count255='140 \advance\count255 by #1\char\count255}
\def\Alphabetic#1{\count255='100 \advance\count255 by #1\char\count255}
\def\Roman#1{\uppercase\expandafter{\romannumeral #1}}
\def\roman#1{\romannumeral #1}
\def\Number#1{\number #1}
\def\BLANC#1{}
\def\titlestyle#1{\par\begingroup \interlinepenalty=9999
     \leftskip=0.02\hsize plus 0.23\hsize minus 0.02\hsize
     \rightskip=\leftskip \parfillskip=0pt
     \hyphenpenalty=9000 \exhyphenpenalty=9000
     \tolerance=9999 \pretolerance=9000
     \spaceskip=0.333em \xspaceskip=0.5em
     \iftwelv@\fourteenpoint\else\twelvepoint\fi
   \noindent #1\par\endgroup }
\def\spacecheck#1{\dimen@=\pagegoal\advance\dimen@ by -\pagetotal
   \ifdim\dimen@<#1 \ifdim\dimen@>0pt \vfil\break \fi\fi}
\def\TableOfContentEntry#1#2#3{\relax}
\def\chapter#1{\par \penalty-300 \vskip\chapterskip
   \spacecheck\chapterminspace
   \chapterreset \titlestyle{\chapterlabel\ #1}
   \TableOfContentEntry c\chapterlabel{#1}
   \nobreak\vskip\headskip \penalty 30000
   \wlog{\string\chapter\space \chapterlabel} }

\def\section#1{\par \ifnum\the\lastpenalty=30000\else
   \penalty-200\vskip\sectionskip \spacecheck\sectionminspace\fi
   \global\advance\sectionnumber by 1
   \xdef\sectionlabel{\the\sectionstyle\the\sectionnumber}
   \wlog{\string\section\space \sectionlabel}
   \TableOfContentEntry s\sectionlabel{#1}
   \noindent {\caps\enspace\sectionlabel\quad #1}\par
   \nobreak\vskip\headskip \penalty 30000 }
\def\subsection#1{\par
   \ifnum\the\lastpenalty=30000\else \penalty-100\smallskip \fi
   \noindent\undertext{#1}\enspace \vadjust{\penalty5000}}

\def\undertext#1{\vtop{\hbox{#1}\kern 1pt \hrule}}
\def\ack{\par\penalty-100\medskip \spacecheck\sectionminspace
   \line{\fourteenrm\hfil ACKNOWLEDGEMENTS\hfil}\nobreak\vskip\headskip }
\def\APPENDIX#1#2{\par\penalty-300\vskip\chapterskip
   \spacecheck\chapterminspace \chapterreset \xdef\chapterlabel{#1}
   \titlestyle{APPENDIX #2} \nobreak\vskip\headskip \penalty 30000
   \TableOfContentEntry a{#1}{#2}
   \wlog{\string\Appendix\ \chapterlabel} }
\def\Appendix#1{\APPENDIX{#1}{#1}}
\def\appendix{\APPENDIX{A}{}}
\def\unnumberedchapters{\let\makechapterlabel=\relax \let\chapterlabel=\relax
   \sectionstyle={\BLANC}\let\sectionlabel=\relax \sequentialequations }
%
%
%
\def\eqname#1{\relax \ifnum\equanumber<0
     \xdef#1{{\noexpand\rm(\number-\equanumber)}}%
       \global\advance\equanumber by -1
    \else \global\advance\equanumber by 1
      \xdef#1{{\noexpand\rm(\chapterlabel\number\equanumber)}} \fi #1}

\def\eqn{\eqno\eqname}

\def\eqinsert#1{\noalign{\dimen@=\prevdepth \nointerlineskip
   \setbox0=\hbox to\displaywidth{\hfil #1}
   \vbox to 0pt{\kern 0.5\baselineskip\hbox{$\!\box0\!$}\vss}
   \prevdepth=\dimen@}}
%

%
%
\def\GENITEM#1;#2{\par \hangafter=0 \hangindent=#1
    \Textindent{$ #2 $}\ignorespaces}
\outer\def\newitem#1=#2;{\gdef#1{\GENITEM #2;}}
\newdimen\itemsize                \itemsize=30pt
\newitem\item=1\itemsize;
\newitem\sitem=1.75\itemsize;     
\newitem\ssitem=2.5\itemsize;     
\outer\def\newlist#1=#2&#3&#4;{\toks0={#2}\toks1={#3}%
   \count255=\escapechar \escapechar=-1
   \alloc@0\list\countdef\insc@unt\listcount     \listcount=0
   \edef#1{\par
      \countdef\listcount=\the\allocationnumber
      \advance\listcount by 1
      \hangafter=0 \hangindent=#4
      \Textindent{\the\toks0{\listcount}\the\toks1}}
   \expandafter\expandafter\expandafter
    \edef\c@t#1{begin}{\par
      \countdef\listcount=\the\allocationnumber \listcount=1
      \hangafter=0 \hangindent=#4
      \Textindent{\the\toks0{\listcount}\the\toks1}}
   \expandafter\expandafter\expandafter
    \edef\c@t#1{con}{\par \hangafter=0 \hangindent=#4 \noindent}
   \escapechar=\count255}
\def\c@t#1#2{\csname\string#1#2\endcsname}
\newlist\point=\Number&.&1.0\itemsize;
\newlist\subpoint=(\alphabetic&)&1.75\itemsize;
\newlist\subsubpoint=(\roman&)&2.5\itemsize;
%

%
%
%
%
\newcount\referencecount     \referencecount=0
\newcount\lastrefsbegincount \lastrefsbegincount=0
\newif\ifreferenceopen       \newwrite\referencewrite
\newif\ifrw@trailer
\newdimen\refindent     \refindent=30pt
\def\NPrefmark#1{\attach{\scriptscriptstyle [ #1 ] }}
\let\PRrefmark=\attach
\def\refmark#1{\relax\ifPhysRev\PRrefmark{#1}\else\NPrefmark{#1}\fi}
\def\refend@{\refmark{\number\referencecount}}
\def\refend{\refend@{}\space }
\def\refsend{\refmark{\count255=\referencecount
   \advance\count255 by-\lastrefsbegincount
   \ifcase\count255 \number\referencecount
   \or \number\lastrefsbegincount,\number\referencecount
   \else \number\lastrefsbegincount-\number\referencecount \fi}\space }
\def\refitem#1{\par \hangafter=0 \hangindent=\refindent \Textindent{#1}}
\def\Ref{\rw@trailertrue\REF}
\def\ref{\Ref\?}

\def\REF#1{\r@fstart{#1}%
   \rw@begin{\the\referencecount.}\rw@end}
\def\REFS#1{\r@fstart{#1}%
   \lastrefsbegincount=\referencecount
   \rw@begin{\the\referencecount.}\rw@end}
\def\r@fstart#1{\chardef\rw@write=\referencewrite \let\rw@ending=\refend@
   \ifreferenceopen \else \global\referenceopentrue
   \immediate\openout\referencewrite=referenc.txa
   \toks0={\catcode`\^^M=10}\immediate\write\rw@write{\the\toks0} \fi
   \global\advance\referencecount by 1 \xdef#1{\the\referencecount}}
{\catcode`\^^M=\active %
 \gdef\rw@begin#1{\immediate\write\rw@write{\noexpand\refitem{#1}}%
   \begingroup \catcode`\^^M=\active \let^^M=\relax}%
 \gdef\rw@end#1{\rw@@end #1^^M\rw@terminate \endgroup%
   \ifrw@trailer\rw@ending\global\rw@trailerfalse\fi }%
 \gdef\rw@@end#1^^M{\toks0={#1}\immediate\write\rw@write{\the\toks0}%
   \futurelet\n@xt\rw@test}%
 \gdef\rw@test{\ifx\n@xt\rw@terminate \let\n@xt=\relax%
       \else \let\n@xt=\rw@@end \fi \n@xt}%
}
\let\rw@ending=\relax
\let\rw@terminate=\relax
\let\splitout=\relax
\def\Closeout#1{\toks0={\catcode`\^^M=5}\immediate\write#1{\the\toks0}%
   \immediate\closeout#1}
%
%
\newcount\figurecount     \figurecount=0
\newcount\tablecount      \tablecount=0
\newif\iffigureopen       \newwrite\figurewrite
\newif\iftableopen        \newwrite\tablewrite
\def\FIG#1{\f@gstart{#1}%
   \rw@begin{\the\figurecount)}\rw@end}

\def\Fig{\rw@trailertrue\def\rw@ending{Fig.~\?}\FIG\?}
\def\fig{\rw@trailertrue\def\rw@ending{fig.~\?}\FIG\?}
\def\TABLE#1{\T@Bstart{#1}%
   \rw@begin{\the\tableecount:}\rw@end}
\def\Table{\rw@trailertrue\def\rw@ending{Table~\?}\TABLE\?}
\def\f@gstart#1{\chardef\rw@write=\figurewrite
   \iffigureopen \else \global\figureopentrue
   \immediate\openout\figurewrite=figures.txa
   \toks0={\catcode`\^^M=10}\immediate\write\rw@write{\the\toks0} \fi
   \global\advance\figurecount by 1 \xdef#1{\the\figurecount}}
\def\T@Bstart#1{\chardef\rw@write=\tablewrite
   \iftableopen \else \global\tableopentrue
   \immediate\openout\tablewrite=tables.txa
   \toks0={\catcode`\^^M=10}\immediate\write\rw@write{\the\toks0} \fi
   \global\advance\tablecount by 1 \xdef#1{\the\tablecount}}
%

%
%
%
\def\getfigure#1{\global\advance\figurecount by 1
   \xdef#1{\the\figurecount}\count255=\escapechar \escapechar=-1
   \edef\n@xt{\noexpand\g@tfigure\csname\string#1Body\endcsname}%
   \escapechar=\count255 \n@xt }
\def\g@tfigure#1#2 {\errhelp=\disabledfigures \let#1=\relax
   \errmessage{\string\getfigure\space disabled}}
\newhelp\disabledfigures{ Empty figure of zero size assumed.}
\def\figinsert#1{\midinsert\Tenpoint\medskip
   \count255=\escapechar \escapechar=-1
   \edef\n@xt{\csname\string#1Body\endcsname}
   \escapechar=\count255 \centerline{\n@xt}
   \bigskip\narrower\narrower
   \noindent{\it Figure}~#1.\quad }
%
%
%
\def\masterreset{\global\pagenumber=1 \global\chapternumber=0
   \global\equanumber=0 \global\sectionnumber=0
   \global\referencecount=0 \global\figurecount=0 \global\tablecount=0 }
\def\FRONTPAGE{\ifvoid255\else\vfill\penalty-20000\fi
      \masterreset\global\frontpagetrue
      \global\lastf@@t=0 \global\footsymbolcount=0}

\def\paperstyle{\letterstylefalse\normalspace\papersize}
\def\letterstyle{\letterstyletrue\singlespace\lettersize}
\def\papersize{\hsize=35 truepc\vsize=50 truepc\hoffset=-2.51688 truepc
               \skip\footins=\bigskipamount}
\def\lettersize{\hsize=5.5 truein\vsize=8.25 truein\hoffset=.4875 truein
	\voffset=.3125 truein
   \skip\footins=\smallskipamount \multiply\skip\footins by 3 }
\paperstyle   
%
%
\def\MEMO{\letterstyle \letterinfo={\hfil } \let\rule=\memorule
	\FRONTPAGE \memohead }
\let\memohead=\relax

\def\memit@m#1{\smallskip \hangafter=0 \hangindent=1in
      \Textindent{\caps #1}}
\def\subject{\memit@m{Subject:}}
\def\topic{\memit@m{Topic:}}
\def\from{\memit@m{From:}}
\def\to{\relax \ifmmode \rightarrow \else \memit@m{To:}\fi }
\def\memorule{\medskip\hrule height 1pt\bigskip}
\newwrite\labelswrite
\newtoks\rw@toks

\def\addressee#1{\null\vskip .5truein\line{
\hskip 0.5\hsize minus 0.5\hsize\the\date\hfil}\bigskip
   \ialign to\hsize{\strut ##\hfil\tabskip 0pt plus \hsize \cr #1\crcr}
   \writelabel{#1}\medskip\par\noindent}
\def\rwl@begin#1\cr{\rw@toks={#1\crcr}\relax
   \immediate\write\labelswrite{\the\rw@toks}\futurelet\n@xt\rwl@next}
\def\rwl@next{\ifx\n@xt\rwl@end \let\n@xt=\relax
      \else \let\n@xt=\rwl@begin \fi \n@xt}
\let\rwl@end=\relax
\def\writelabel#1{\immediate\write\labelswrite{\noexpand\labelbegin}
     \rwl@begin #1\cr\rwl@end
     \immediate\write\labelswrite{\noexpand\labelend}}
\newbox\FromLabelBox

\let\labelend=\relax
\def\labelbegin#1\labelend{\setbox0=\vbox{\ialign{##\hfil\cr #1\crcr}}
     \MakeALabel }
\newtoks\FromAddress
\FromAddress={}
\def\MakeFromBox#1{\global\setbox\FromLabelBox=\vbox{\Tenpoint
     \ialign{##\hfil\cr #1\the\FromAddress\crcr}}}
\newdimen\labelwidth		\labelwidth=6in
\def\MakeALabel{\vskip 1pt \hbox{\vrule \vbox{
	\hsize=\labelwidth \hrule\bigskip
	\leftline{\hskip 1\parindent \copy\FromLabelBox}\bigskip
	\centerline{\hfil \box0 } \bigskip \hrule
	}\vrule } \vskip 1pt plus 1fil }
\newskip\signatureskip       \signatureskip=30pt
\def\signed#1{\par \penalty 9000 \medskip \dt@pfalse
  \everycr={\noalign{\ifdt@p\vskip\signatureskip\global\dt@pfalse\fi}}
  \setbox0=\vbox{\singlespace \ialign{\strut ##\hfil\crcr
   \noalign{\global\dt@ptrue}#1\crcr}}
  \line{\hskip 0.5\hsize minus 0.5\hsize \box0\hfil} \medskip }
\newbox\letterb@x
\def\lettertext{\par\unvcopy\letterb@x\par}
\def\multiletter{\setbox\letterb@x=\vbox\bgroup
      \everypar{\vrule height 1\baselineskip depth 0pt width 0pt }
      \singlespace \topskip=\baselineskip }
\def\letterend{\par\egroup}
%
%
%
\newskip\frontpageskip
\newtoks\Pubnum
\newtoks\pubtype
\newif\ifp@bblock  \p@bblocktrue
\def\PH@SR@V{\doubl@true \baselineskip=24.1pt plus 0.2pt minus 0.1pt
             \parskip= 3pt plus 2pt minus 1pt }
\def\PHYSREV{\paperstyle\PhysRevtrue\PH@SR@V}
\def\titlepage{\FRONTPAGE\paperstyle\ifPhysRev\PH@SR@V\fi
   \ifp@bblock\p@bblock \else\hrule height\z@ \relax \fi }
\def\nopubblock{\p@bblockfalse}

\frontpageskip=12pt plus .5fil minus 2pt
\pubtype={\tensl Preliminary Version}
\Pubnum={}
\def\p@bblock{\begingroup \tabskip=\hsize minus \hsize
   \baselineskip=1.5\ht\strutbox \topspace-2\baselineskip
   \halign to\hsize{\strut ##\hfil\tabskip=0pt\crcr
       \the\Pubnum\crcr\the\date\crcr\the\pubtype\crcr}\endgroup}
\def\title#1{\vskip\frontpageskip \titlestyle{#1} \vskip\headskip }
\def\author#1{\vskip\frontpageskip\titlestyle{\twelvecp #1}\nobreak}

\def\address#1{\par\kern 5pt\titlestyle{\twelvepoint\it #1}}
\def\andaddress{\par\kern 5pt \centerline{\sl and} \address}

\def\abstract{\par\dimen@=\prevdepth \hrule height\z@ \prevdepth=\dimen@
   \vskip\frontpageskip\centerline{\fourteenrm ABSTRACT}\vskip\headskip }

%
%
%

\def\\{\relax \ifmmode \backslash \else {\tt\char`\\}\fi }
\def\sequentialequations{\relax\if\equanumber<0\else\global\equanumber=-1\fi}

\def\journal#1&#2(#3){\unskip, \sl #1\unskip~\bf\ignorespaces #2\rm (19#3),}

\def\topspace{\hrule height 0pt depth 0pt \vskip}

\def\Buildrel#1\under#2{\mathrel{\mathop{#2}\limits_{#1}}}
\def\becomes#1{\mathchoice{\becomes@\scriptstyle{#1}}{\becomes@\scriptstyle
   {#1}}{\becomes@\scriptscriptstyle{#1}}{\becomes@\scriptscriptstyle{#1}}}
\def\becomes@#1#2{\mathrel{\setbox0=\hbox{$\m@th #1{\,#2\,}$}%
	\mathop{\hbox to \wd0 {\rightarrowfill}}\limits_{#2}}}

\let\int=\intop         
\def\lsim{\mathrel{\mathpalette\@versim<}}
\def\gsim{\mathrel{\mathpalette\@versim>}}
\def\@versim#1#2{\vcenter{\offinterlineskip
	\ialign{$\m@th#1\hfil##\hfil$\crcr#2\crcr\sim\crcr } }}
\def\big#1{{\hbox{$\left#1\vbox to 0.85\b@gheight{}\right.\n@space$}}}
\def\Big#1{{\hbox{$\left#1\vbox to 1.15\b@gheight{}\right.\n@space$}}}
\def\bigg#1{{\hbox{$\left#1\vbox to 1.45\b@gheight{}\right.\n@space$}}}
\def\Bigg#1{{\hbox{$\left#1\vbox to 1.75\b@gheight{}\right.\n@space$}}}
%
%
%
\let\sec@nt=\sec
\def\sec{\relax\ifmmode\let\n@xt=\sec@nt\else\let\n@xt\section\fi\n@xt}
\def\obsolete#1{\message{Macro \string #1 is obsolete.}}
\def\firstsec#1{\obsolete\firstsec \section{#1}}
\def\firstsubsec#1{\obsolete\firstsubsec \subsection{#1}}
\def\thispage#1{\obsolete\thispage \global\pagenumber=#1\frontpagefalse}
\def\thischapter#1{\obsolete\thischapter \global\chapternumber=#1}
\def\REFSCON{\obsolete\REFSCON\REF}
\def\splitout{\obsolete\splitout\relax}
\def\prop{\obsolete\prop \propto }
\def\nextequation#1{\obsolete\nextequation \global\equanumber=#1
   \ifnum\the\equanumber>0 \global\advance\equanumber by 1 \fi}
\def\BOXITEM{\afterassigment\B@XITEM\setbox0=}
\def\B@XITEM{\par\hangindent\wd0 \noindent\box0 }
\def\phyzzx{PHY\setbox0=\hbox{Z}\copy0 \kern-0.5\wd0 \box0 X}
%
%
\everyjob{\xdef\today{\monthname\ \number\day, \number\year}}
        
%


\hoffset=0.2truein
\voffset=0.1truein
\hsize=6truein

\def\CALT#1{\hbox to\hsize{\tenpoint \baselineskip=12pt
	\hfil\vtop{\hbox{\strut CALT-68-#1}
	\hbox{\strut DOE RESEARCH AND}
	\hbox{\strut DEVELOPMENT REPORT}}}}

\def\CALTECH{\smallskip
	\address{California Institute of Technology, Pasadena, CA 91125}}
\def\TITLE#1{\vskip 1in \centerline{\fourteenpoint #1}}
\def\AUTHOR#1{\vskip .5in \centerline{#1}}

\def\ABSTRACT#1{\vskip .5in \vfil \centerline{\twelvepoint \bf Abstract}
	#1 \vfil}

\def\sqr#1#2{{\vcenter{\hrule height.#2pt
      \hbox{\vrule width.#2pt height#1pt \kern#1pt
        \vrule width.#2pt}
      \hrule height.#2pt}}}

\def\section#1#2{
\noindent\hbox{\hbox{\bf #1}\hskip 10pt\vtop{\hsize=5in
\baselineskip=12pt \noindent \bf #2 \hfil}\hfil}
\medskip}

\def\underwig#1{	
	\setbox0=\hbox{\rm \strut}
	\hbox to 0pt{$#1$\hss} \lower \ht0 \hbox{\rm \char'176}}

\def\bunderwig#1{	
	\setbox0=\hbox{\rm \strut}
	\hbox to 1.5pt{$#1$\hss} \lower 12.8pt
	 \hbox{\seventeenrm \char'176}\hbox to 2pt{\hfil}}

\def\MEMO#1#2#3#4#5{
\frontpagetrue
\centerline{\tencp INTEROFFICE MEMORANDUM}
\smallskip
\centerline{\bf CALIFORNIA INSTITUTE OF TECHNOLOGY}
\bigskip
\vtop{\tenpoint \hbox to\hsize{\strut \hbox to .75in{\caps to:\hfil}
\hbox to3in{#1\hfil}
\hbox to .75in{\caps date:\hfil}\quad \the\date\hfil}
\hbox to\hsize{\strut \hbox to.75in{\caps from:\hfil}\hbox to 2in{#2\hfil}
\hbox{{\caps extension:}\quad#3\qquad{\caps mail code:\quad}#4}\hfil}
\hbox{\hbox to.75in{\caps subject:\hfil}\vtop{\parindent=0pt
\hsize=3.5in #5\hfil}}
\hbox{\strut\hfil}}}


\def\section#1{\par\ifnum\the\lastpenalty=30000\else
        \penalty-200\vskip\sectionskip\spacecheck\sectionminspace\fi
        \global\advance\sectionnumber by 1
        \xdef\sectionlabel{\the\sectionstyle\the\sectionnumber}
        \wlog{\string\section\space\sectionlabel}
        \TableOfContentEntry s\sectionlabel{#1}
        \noindent {\caps\enspace\sectionlabel\quad #1}\par
        \nobreak\vskip\headskip\penalty 30000 }

\def\oldeq#1{{\let\chapterlabel=\relax #1}}

\def\d{\vec{\rm \bf d}}
\def\l{\vec{\ell}}\def\d{\vec{\rm \bf d}}
\def\l{\vec{\ell}}
\def\d{\vec{\rm \bf d}}
\def\l{\vec{\ell}}
\def\p{\partial\kern-.614em\raise.65ex\hbox{${}^\leftrightarrow$}\kern-.272em}

\def\r{\vec{r}}
\def\half{{1\over 2}}
\def\del{\vec{\nabla}}


\CALT{1865}   
\TITLE{A Global Analog of Cheshire Charge\footnote*{Work supported in
part by U.S. Department of Energy Grant no.  DE-FG03-92-EP40701}}
\AUTHOR{Patrick McGraw\footnote{\dag}{pmcgraw@theory.caltech.edu}}
\CALTECH

\ABSTRACT{It is shown that a model with a spontaneously broken global symmetry
can support defects analogous to Alice strings, and a process analogous to
Cheshire charge exchange can take place.  A possible realization in superfluid
He-3 is pointed out. }

\section{Introduction}
March-Russell, Preskill, and Wilczek \Ref\MPW{J.~March-Russell, J.~Preskill,
F.~Wilczek.  Phys.~Rev.~Lett. {\bf 68} (1992), 2567.} showed that vortices of a
theory with a global $U(1)$ symmetry broken to $Z_2$ can scatter quanta of
$Z_2$ charge with a cross section almost equal to to the maximal Aharonov-Bohm
cross-section, due to a frame-dragging of local mass eigenstates. Here we
demonstrate the realization of a non-abelian Aharonov-Bohm phenomenon (Cheshire
charge) in the context of a global model.

In the first section, we describe a relativistic field theory that supports a
global analog of Alice strings
\Ref\Alice{A.~Schwarz, Nuc.~Phys.~{\bf B208} (1982) 141.}
and then describe how the process of charge exchange occurs by means of quantum
interference.  Some differences as well as similarities to the parallel
phenomenon of Cheshire charge in gauge theories are mentioned, as well as the
manner in which the global phenomenon can be viewed as a limit of the gauge
case at very weak gauge coupling.

One of the motivations for studying global vortices and global Aharonov-Bohm
scattering is that many more global than local symmetry-breaking transitions
are available for manipulation in condensed matter systems.  The possibility
arises of finding condensed-matter systems which can serve as laboratory
analogs of otherwise observationally inaccessible gauge string phenomena.
In section 2, we consider the possibility of finding  a laboratory analog of
Cheshire charge in the superfluid A phase of helium-3.  The  group-theoretic
properties necessary for the existence of Cheshire charge are present in He-3
A,  although in practice it may be difficult to devise an experiment to observe
it.

\section{Cheshire Charge in a Theory With Broken Global Symmetry}
\subsection{The Model}

Consider a theory with a global $SO(3)$ symmetry containing a Higgs scalar
$\Phi$ transforming as the 5-dimensional symmetric tensor representation, which
we will write as a $3\times 3$ $SO(3)$ matrix, and another scalar field $\Psi$
transforming as a 3-dimensional vector.  $\Psi$ will serve as the test particle
that scatters from vortices. The fields transform under $SO(3)$ according to:
$$\Psi \rightarrow \Omega \Psi,\ \  \Phi \rightarrow \Omega \Phi
\Omega^{-1}.\eqn\a$$
where $\Omega$ is $3\times 3$ $SO(3)$ matrix. We will denote the generators of
$G=SO(3)$ as:
$$ T_{1}= \left[ \matrix{
                  0 & 0 & 0 \cr
		  0 & 0 & -i \cr
		  0 & i & 0
		 } \right],\
 T_{2}= \left[ \matrix{
                  0 & 0 & i \cr
		  0 & 0 & 0 \cr
		  -i & 0 & 0
		 } \right],\
 T_{3}= \left[ \matrix{
                  0 & -i & 0 \cr
		  i & 0 & 0 \cr
		  0 & 0 & 0
		 } \right].\eqn\a$$
We also introduce a bilinear coupling of the $\Psi$ fields to the Higgs:
$$\Delta{\cal L}=\lambda\Psi^{T}\Phi\Psi.\eqn\a $$
Now let the Higgs field acquire a vacuum expectation value
$\Phi_{0}=\nu diag(1,1,-2)$.
This breaks the symmetry group down to $H=U(1)\times_{S.D.}Z_{2}$. This is the
same symmetry breaking pattern previously considered in the case of Alice
strings [\Alice] ; the difference is that we are considering a global, rather
than a gauge, symmetry.  The VEV $\Phi_{0}$ induces a mass splitting among the
the members of the multiplet $\Psi$, much as in Reference [\MPW]. The first two
components of $\Psi$ are degenerate and are mixed by the unbroken $U(1)$
generator $T_{3}$, while the third component is an H singlet. From the first
two we can form basis eigenstates of opposite $U(1)$ charges:
$$ u_{+}=(1,i,0),\  u_{-}=(1,-i,0).\eqn\a$$

The VEV in this theory can be thought of as taking values on the surface of a
sphere with antipodal points identified.  A visual analogy for the
symmetry-breaking pattern is the director field of a nematic liquid crystal
(NLC). The order parameter of the NLC, like that of our theory, can be thought
of as an undirected line segment at each point in space. The group of
transformations which leave this segment invariant include continuous rotations
about the director's axis (the $U(1)$ component) as well as a discrete
$180^{\circ}$
rotation about an axis perpendicular to the segment.  This $180^{\circ}$ flip
generates the $Z_{2}$ component.  Since the discrete $180^{\circ}$ rotation
does not commute with continuous rotations about the preferred axis, the full
unbroken group is a {\it semidirect} product.  This visual analogy will be
useful later in explorations of condensed-matter systems.  Our model can be
reformulated in terms of a director field $\d$, a vector in internal space,
rather than a tensor, by defining
$$ \Phi_{ab} = d_{a}d_{b} - d^{2}\delta_{ab} \eqn\a$$
and making the identification $\d\equiv -\d$.

\subsection{Construction of Alice Strings and Vortices}

The above model can form topologically stable $\pi_{1}$ type defects. For
simplicity we consider the model in two spatial dimensions, so that the defects
are vortices. All arguments can be generalized straightforwardly to strings in
three spatial dimensions. A vortex with core at the origin could have an
asymptotic field configuration far from the origin given in polar coordinates
by:
$$\Phi({\bf x})=\exp [{{-i\varphi T_{1}}\over {2}}]\Phi_{0}\exp [{{i\varphi
T_{1}}\over {2}}].\eqn\a$$
The Higgs field is single valued, but the mass eigenstates of the $\Psi$ field
are not well-defined globally. One can define local (frame-dragged) mass
eigenstates $\rho_{i}$ at any point outside the core by $\rho=\exp[{-i\varphi
T_{1}\over {2}}]\Psi$.


As in reference [\MPW], these local eigenstates define a frame at each point
outside the core, and a state is adiabatically transported if its components in
the local basis remain unchanged at each point. Notice that
$\rho_{2}(0)=-\rho_{2}(2\pi).$ This means that when a state $\rho_{2}$ is
adiabatically transported through a loop that encloses the vortex core once, it
acquires an Aharonov-Bohm-like phase of $-1$, whereas $\rho_{1}$ acquires no
such phase.
In terms of the fields of definite U(1) charge,  $\rho_{+} = \rho_1 + i \rho_2$
and $\rho_{-} = \rho_1 - i \rho_2$,  the boundary condition can be written as
$\rho_{+}(2 \pi ) = \rho_{-}(0),\/ \rho_{-}(2 \pi ) = \rho_{+}(0)$.
This means that the sign of the global U(1) charge is reversed when a $\rho$
particle is adiabatically transported once around the string.
Thus we have an global analog of the Alice string that occurs in the
corresponding gauge theory.

In the gauge theory case it is known that a vortex-antivortex pair can acquire
a ``Cheshire charge'' which compensates the charge gained by the particle upon
threading the pair, so that the total charge is actually conserved.
\Ref\Cheshire{M.~Alford,  K.~Benson, S.~Coleman,  J.~March-Russell, F.~Wilczek.
 Phys. Rev. Lett. {\bf 64} (1990), 1632. \/ J.~Preskill,  L.~Krauss.  Nuc Phys
{\bf B 341} (1990), 50.} This charge is a global property of the vortex pair;
it cannot gauge-invariantly be localized to either one of the vortices or to
any region of space near them.  To see how charge conservation is maintained in
this global model, consider a vortex-antivortex pair with the two cores
separated along the x-axis by a distance D which is large compared to the core
radius.  Outside the cores, the vortex-antivortex solution to the field
equations can be written:
$$\Phi_{0}({\bf x})=\exp[{-i\Delta\varphi T_{1}\over
{2}}]\Phi_{0}\exp[{i\Delta\varphi T_{1}\over {2}}]\eqn\phizero$$
where $\Phi_{0}=\nu \rm diag(1,1,-2)$ and
$\Delta\varphi=\varphi_{1}-\varphi_{2}$ as defined in  figure
\FIG\DeltaTheta{Definition of angles $\varphi_1$, $\varphi_2$, and
$\Delta\varphi$
in vortex pair geometry.}\DeltaTheta.
(Since the fluxes of the two vortices lie in the same
$U(1)$ subgroup,  it is easy to show that when the Higgs field is constrained
to lie in the vacuum manifold,  the static 2-d field equations reduce to
Laplace's
equation.  Thus $\del\Phi$ is dual to the electric field of two opposite
charges in two dimensions,  and the solution \phizero\/ is obtained.)

At points far away from both cores $(r\gg D)$, the Higgs field approaches a
single asymptotic value $\Phi_{0}$. Thus the embedding of the unbroken group H
is well-defined on any large circle outside the the two cores: its connected
component is the $U(1)$ generated by $T_{3}$.

 The local mass eigenstates are given by $\rho=\exp[{-i\Delta\varphi
T_{1}/2}]\Psi$ and are thus unchanged under adiabatic transport along paths
that remain far away from the pair
 where $\Delta\varphi\approx 0$.  However, if a $\Psi_{2}$ state is transported
 from $y=-\infty$ to $y=+\infty$ along the
 y-axis (or along any path that passes between the two cores),
 the angle $\Delta\varphi$ winds through $2\pi$,
 and the state $\Psi_{2}$ acquires an Aharonov-Bohm phase of -1,
 whereas $\Psi_{1}$ acquires no such phase.
 None of the triplet components acquires a phase if it is transported from
$y=-\infty$ to $y=+\infty$ on a path which does not pass between the two cores.
Thus an eigenstate of $U(1)$ charge changes the sign of its charge when it
passes through the pair.

In order to understand how charge conservation is maintained, we must
realize that there is an infinite family of vortex pair solutions related by
$U(1)$ rotations.  The solutions
$$\Phi_{\alpha}({\bf x})=\exp(i\alpha T_{3})\Phi_{0}({\bf x})\exp(-i\alpha
T_{3}),\eqn\alphastates$$
where $0<\alpha<2\pi$ and $\Phi_{0}({\bf x})$ is given by \phizero,
 all have the same energy because they are related by a global
 symmetry transformation. Since they all have the same asymptotic   value of
the Higgs field, they can be continuously deformed into each other;
thus there is a charge rotor zero mode. Figure
\FIG\DegStates { Degenerate configurations of vortex pair.  This figure shows
              two order parameter configurations for a vortex-antivortex pair,
              related by a global symmetry operation.  The order parameter
              is represented by an undirected line segment.  In Fig.2A, the
              directors all lie within the plane of the page.  In Fig.2B,  they
              rotate outward, and only their projection in the plane of the
              page is shown.}\DegStates\/
shows the action of the symmetry on the field of a vortex-antivortex pair.
The order parameter is drawn as an undirected line segment as discussed above.
Figure 2a shows one representative of a class of flux eigenstates.
 Other states degenerate with this one are obtained by rotating each of the
directors through arbitrary angle $\alpha$ about the x-axis: Fig. 2b shows
the result when $\alpha=\pi/2$.
(Note that unlike our model, {\it physical}\/ NLC's do not in general possess a
continuous degeneracy of this type, but only a twofold degeneracy, because the
free energy is not invariant under
purely internal rotations of the director, but only under rotations of the
whole coordinate frame.\Ref\NLC{See, for example, I.~Chuang, R.~Durrer,
N.~Turok, B.~Yurke.  Science {\bf 251} (1991), 1336.}  The broken symmetry in
NLC's is not truly an internal one of the type that occurs in relativistic
field theories.)

  The pair states which transform as irreducible representations of the
asymptotically unbroken $U(1)$ group are coherent superpositions of the
solutions \alphastates:  they are the quantized energy levels of the zero mode.
 A state
with charge $n$ is given by:

$$|n> = \int_0^{2\pi} d\alpha\exp(-in\alpha )|\Phi_{\alpha}>.\eqn\a$$

\subsection{The Charge Transfer Process}
To understand the process of charge transfer by which a test particle reverses
its charge and the vortex pair acquires a compensating charge, consider first
the case where the vortex pair is originally in the state $|\alpha >$
 described
above.
For each value of $\alpha$, there is one component
of $\Psi$ that
acquires a phase upon passing between the vortices,
and another which does not.
However, these states depend on the value of $\alpha$.
 Let $u_{1\alpha}$ be
the state which acquires no phase, and $u_{2\alpha}$ be the state which
acquires a phase of $-1$. Then:
$$u_{1\alpha}=(\cos\alpha,-\sin\alpha,0), \
u_{2\alpha}=(\sin\alpha,\cos\alpha,0).\eqn\a$$
The $U(1)$ charge eigenstates $u_{+}=(1,i,0)$ and $u_{-}=(1,-i,0)$ are
expressed in terms of $u_{1\alpha}$ and $u_{2\alpha}$ as:
$$u_{+}=\exp(-i\alpha)(u_{1\alpha}+i u_{2\alpha}),\
  u_{-}=\exp(i\alpha)(u_{1\alpha}-i u_{2\alpha}).\eqn\a$$
Thus,  when the state $u_{+}$ is adiabatically transported along a path that
threads the pair,
it is turned into the state $\exp (-2i\alpha)u_{-}$, whereas $u_{-}$ becomes
$\exp (2i\alpha)u_{+}$. These relations may be expressed simply in terms of
a monodromy matrix, written in the charge eigenstate basis as follows:
$$\rho(2\pi) = {\cal M}(\alpha) \rho(0),$$
where
$${\cal M}(\alpha) = \left[ \matrix{ 0  &e^{2i\alpha} \cr
                                     e^{-2i\alpha} &0} \right].
                     \eqn\monodromy$$

Now take
an initial state in which the vortex pair is in the charge-zero
eigenstate
$|0>$ and the test particle is in the state $u_{+}$:
$$|u_{+}>\otimes|0>=\int_0^{2\pi} d\alpha
|u_{+}>\otimes|\Phi_{\alpha}>,\eqn\a$$
After  the particle is dragged through the loop, the final state will be:
$$\int_0^{2\pi} d\alpha
\exp(-2i\alpha)|u_{-}>\otimes|\Phi_{\alpha}>=|u_{-}>\otimes
|2>.\eqn\finalstate$$
The state has evolved into one in which the vortex pair has charge $+2$,
because of the different phases acquired by the wavefunction
in the different
$\alpha$ sectors.  The zero mode has been excited by means of a
quantum-mechanical interference process,
 which is the usual means for the
transfer of Cheshire charge, except that in this case it has occurred in a
model
with no gauge symmetry and does not have any topological interpretation
\Ref\Bucher{M.~Bucher, H.K.~Lo, J.~Preskill. Nuc. Phys. B 386 (1992), 3.} in
terms of
lines of electric flux being trapped between the vortices.

It may be noticed that even though the charged states $|n>$ exist for all
integers, only those with $n$ even can be produced by this process from an
initially uncharged vortex. This is not necessarily the case in gauge Alice
models.  We can, in the case of a gauge model, take the initial gauge group
$G$ to be $SU(2)$ rather than $SO(3)$, and include matter fields transforming
in the spinor representation.  After the symmetry is broken to
$U(1)\times_{S.D.} Z_2$, the spinor components become two oppositely charged
states which interchange under the action of the $Z_2$ flip.  The smallest
electric charge in the theory is that carried by these spinor particles, and
it is by passing these  through a loop of Alice string (or pair of
Alice vortices) that the
odd-numbered Cheshire charge states are excited.
However, the frame-dragging effect considered in Ref. [\MPW] and in the present
paper requires a matter field bilinearly coupled to the symmetry-breaking order
parameter.  Since the Higgs field in our model transforms in the 5-dimensional
representation,  no singlet can be formed from
the Higgs field with only two spinors, and we are forced to consider matter
fields $\Psi$ lying in a vector representation.  Thus, in comparing our global
Alice system to the corresponding gauge model, the states which we have
called $\rho_{\pm}$  should be thought of as doubly charged.
The monodromy matrix \monodromy, for example, has the property
${\cal M}^2 = 1$, rather than ${\cal M}^2 = -1$ as in the case of singly
charged
objects.\Ref\PreskillLo{H.K.~Lo, J.~Preskill,  Phys. Rev. D {\bf  48} (1993)
4821.}

\subsection{Comparison With Ordinary Gauge Cheshire Charge}

It is interesting that, although the existence of Cheshire charge is a
consequence of the symmetry breaking pattern only, and occurs in this global
model for the same reason as in a gauge theory, the nature of the charge is
rather different.  As we see in the following paragraphs,  global Cheshire
charge actually is localizable and is carried by the scalar fields rather than
the vector fields.

In the global theory we are considering, the zero mode  is simply a ``rigid''
$U(1)$ rotation of the entire Higgs field configuration:  thus it is a subgroup
of the global $SO(3)$ transformations. $\alpha$ is the coordinate of the zero
mode, and {\it classically\/} the excitations of this mode are
states where
$d\alpha/dt \not= 0$.  Quantum mechanically, $d\alpha/dt$ will be replaced by a
canonical momentum with discrete eigenvalues.
This is to be contrasted with the
case of a gauge model, where  a  rigid rotation of the Higgs fields alone
fails to satisfy the equations of motion.  In temporal gauge,  the zero mode
can be written \Ref\ZM{M.~Alford, K.~Benson, S.~Coleman,J.~March-Russell,
F.~Wilczek. Nuc.~Phys~B {\bf 349} (1991) 414.}:
$$\Phi=\Omega\bar{\Phi}\Omega^{-1}, \
 A_{\mu}=\Omega \bar{A}_{\mu}\Omega^{-1} +
           \delta_{\mu}^{i}\partial_{i}\Omega \Omega^{-1},\eqn\a$$
where $\bar{\Phi}$ and $\bar{A}_{\mu}$ are static solutions and $\Omega(x,t)$
is
a {\it spatially varying}
 $SO(3)$ transformation that tends to an element of the unbroken group H (
namely $\exp(i\alpha T_{3})$) at infinity.
 (Notice that if the term $\delta_{\mu}^{i}\partial_{i}\Omega \Omega^{-1}$ were
replaced by $\partial_{\mu}\Omega \Omega^{-1}$ then this would be a
physically irrelevant gauge transformation.)  By transforming to another gauge,
one can view the zero mode as purely an excitation of the gauge fields,
$$ \Phi = \bar{\Phi}, A_{\mu}=\bar{A}_{\mu}-
   \delta_{\mu}^{0}\partial_{0}\Omega^{-1} \Omega,\eqn\a$$
whereas in the global case it is purely an excitation of the scalar fields.

In our global case one can see that there is a nonzero charge density (i.e. ,
0 component of the global current) which is localized in the region of space
surrounding the vortices:
$$J_{0(3)}= 2{\rm Tr}\Phi T_{3}\p_{0}\Phi =
36\nu\sin^2(\Delta\varphi/2)(d\alpha/d t).\eqn\a$$
This definitely localizable charge carried by the scalar fields contrasts with
the usual case, where the charge is carried by the gauge
fields and its
apparent location can be moved by performing gauge
 transformations
\Ref\Everett{A.~Everett, Phys.~Rev.~D {\bf 47} (1993), R1227.}.
Since the global charge density is locally measurable, it must be the case that
the charge is gradually transferred as the the charged particle moves between
the vortex cores.  One must suppose that the charge density propagates away
from the particle as the particle moves,  and spreads itself through the region
of space surrounding the cores.  The transfer of gauge Cheshire charge, on the
other hand,  cannot gauge-invariantly be said to happen at a particular time
and place, or even incrementally at a well-defined rate.

The relation between charge and energy is also different in the global case
compared to the  gauge case. This can be demonstrated by quantizing the zero
mode.  First treat $\alpha$ as
a classical coordinate.  Assume that the Higgs field configuration is given by
$$ \Phi(x,t)= \exp[i\alpha(t)T_{3}]\Phi_{0}(x)\exp[-i\alpha(t)T_{3}]\eqn\a$$
i.e.,  a time-varying rigid rotation of the angle $\alpha$.  Then
$ \partial_{t}\Phi = i\dot{\alpha}[T_{3},\Phi_{0}] $
and the gradient part of the Lagrangian gives
$$ \  L = \int dt d^{3}x  \partial_{0}\Phi\cdot\partial^{0}\Phi
        = -\int dt {\int d^{3}x{\rm Tr}([T_{3},\Phi_{0}]}\dot{\alpha})^2,
	 \eqn\a$$
(Since the static configurations of different $\alpha$ are degenerate, all
other terms in the Lagrangian are independent of $\alpha$ and do not enter in
that coordinate's equations of motion.) Letting
$$ I = \int  d^{3}x{\rm Tr}[T_{3},\Phi_{0}]^2,\eqn\a$$
we can define a momentum $\Pi_{\alpha}$ conjugate to $\alpha$ and write a
Hamiltonian
$$ H = {\Pi_{\alpha}^{2}\over {2I}}.\eqn\ZMHam$$
Now the commutator $[T_3, \Phi_{0}]$  is of order $|\Phi|$ within some volume
surrounding the vortex cores.  Since there is no domain wall connecting the
cores, $D$ is the only relevant length scale, and dimensional analysis then
dictates that the ``moment of inertia'' $I$ scales as $D^d$, where $d$ is the
number of spatial dimensions.
Accordingly, the charge rotor Hamiltonian has eigenvalues
 $E_{n}= {n^{2}\hbar^{2}\over{2I}}\approx n^2 \hbar^2 /(|\Phi|^2\times{\rm
Volume})$,
 so the energy splittings among the global charge eigenstates scale as
$1/D^{d}$.
One may contrast this with
the Coulomb energy  (logarithmic in 2 space dimensions, $1/D$ in 3)
of the usual gauge Cheshire charge.

The global vortex pair can be considered
as a limit of the gauge model
where
the gauge coupling is so small that the
Compton wavelength of
the massive
vector bosons is much larger than the
separation $D$ of the two vortex centers.
In this limit, the Higgs field cores
(regions where the Higgs leaves the vacuum
manifold) can remain small while the gauge
field cores (regions of nonzero
magnetic flux) of the vortices become much larger than $D$
so that there is
actually no winding of the gauge field near the pair.
 In this case all the
fields are defined on a trivial bundle
and the Higgs VEV is
covariantly non-constant just as in the global model.
 Presumably the Cheshire
charge states will behave as in the global case,
with a gauge invariantly
localizable charge density carried by
the scalars near the string.

\subsection{Scattering From a Global Alice String}

It is worth noting briefly that a single global Alice string of the type we
have constructed will scatter incoming quanta of the $\Psi$ fields.
In fact,  if we limit our attention to the $\rho_{2}$ and $\rho_3$ components,
the calculation of the scattering amplitude proceeds
precisely as in reference [\MPW].  $\rho_{1}$, on the other hand, does not
scatter at all (except perhaps off the vortex core itself).  In other words,
$\rho_1$ and $\rho_2$ are the monodromy eigenstates, and $\rho_1$ has
eigenvalue unity.  The result is that,  if an incoming plane wave consists of
either of the charge eigenstates $\rho_+$  and $\rho_-$,  the scattered wave
will be pure $\rho_2$,  which is a superposition $-i(\rho_+ - \rho_-)/2$  of
the two charge eigenstates.  Consider the scattering of
a $\rho_+$ incident at momentum below the threshold for $\rho_3$ production.
We expect the $\rho_1$ and $\rho_2$  to behave asymptotically as an incoming
and a scattered wave.  We may write this asymptotic behavior as follows:
$$\left( \matrix{\rho_1 \cr \rho_2}\right) =
  \left( \matrix{1 & 0 \cr 0 & e^{i\varphi/2}}\right)
  \left\{ {
  \left( \matrix{1 \cr i}\right) e^{-ikx} +
  \left( \matrix{1 \cr i}\right) f_{+}(\varphi){e^{ikr}\over r^{1/2}}+
  \left( \matrix{1 \cr -i}\right) f_{-}(\varphi){e^{ikr}\over r^{1/2}}
   }\right\}.\eqn\a$$
$f_+$ and $f_-$ are the charge-preserving and charge-reversing amplitudes
for the scattered particle.  The diagonal matrix in front enforces the boundary
conditions on the frame-dragged states.  For simplicity we are considering a
vortex in the flux eigenstate with $\alpha = 0$.  Proceeding by analogy with
[\MPW],  the equations
of motion for the second component lead to
$$f_{+}(\varphi)-f_{-}(\varphi)={e^{-i\varphi/2}\over{(2\pi i k)^{1/2}}}
  \left({1\over{\cos(\varphi/2)}}+2\sum_{n=0}^\infty (-1)^n(e^{i\Delta_n}-1)
         \cos\left[ (n+{1\over2})\varphi \right]\right),$$
$$\Delta_n = \pi \left( n + \half - \sqrt{(n+\half)^2+{1\over 4}}
             \right).\eqn\amplitudes.$$
Since the first component does not scatter,  we also have
$f_{+}(\varphi)+f_{-}(\varphi)=0$ and the amplitudes are uniquely determined.
The differential cross-sections for charge-preserving and charge-flipped
scattering,
$d\sigma_+/d\theta$\/ and $d\sigma_-/d\theta$,  are identical to each other and
equal to one fourth the
scattering cross-section derived  in reference [\MPW] for an abelian
global vortex:
$${d\sigma_+ \over d\theta} = {d\sigma_- \over d\theta} =
{1\over 8\pi k}{1\over\sin^2(\theta/2)}(1+C(\theta)) \eqn\crosssec$$
where $\theta = \pi -\varphi$ is the scattering angle and $C(\theta)$, which
vanishes at $\theta = 0$, is a function obtained by summing (and squaring) the
series in \amplitudes.
The inclusive cross section $d\sigma_+/d\theta + \sigma_-/d\theta$\/ is half
that of the abelian
case considered in [\MPW], because the $\rho_1$ state is, so to speak,
``filtered out''  of the scattered wave, just as a linear polarizer halves the
intensity of a circularly polarized light beam.

Setting $C(\theta)=0$ in \crosssec\/
would give the cross section for doubly charged projectiles scattering from
a gauge Alice string.
$C(\theta)$, a correction present only in the global analog, results from
diagonal $1/r^2$ potential
terms appearing in the equations of motion for the $\rho$ fields.\footnote*{
This cross-section was derived by neglecting  off-diagonal terms that cause
mixing of $\rho_2$ and $\rho_3$ near
 the vortex.  Navin's analysis \Ref\Navinthesis{ R. Navin, {\it Global
Analogue of the Aharonov-Bohm Effect}, Ph.D. Thesis (Caltech 1993) and Caltech
preprint CALT-68-1896}
suggests that the corrections to the standard Aharonov-Bohm cross section may
disappear when the scattering problem is
solved exactly.}
Evidently, these corrections modify the inclusive cross section but they do not
affect the ratio $\sigma_+/\sigma_-$, which depends on the monodromy properties
of the scattered particles and not on the local or global nature of the
vortices.

\section{A Condensed-Matter Example}

Cheshire charge is a generic phenomenon that occurs when a theory has vortices
whose winding fails to commute with a generator of the unbroken symmetry group
of the vacuum.  The results of the previous section show that Cheshire charge
also arises if the theory has only global and not gauge symmetries.  The
essential group-theoretic concept is the same although the mechanism is
different.

In this section, we discuss a physical system which exhibits the type of
symmetry-breaking necessary for the existence of Cheshire charge:  one which
allows mixing of flux eigenstates within a conjugacy class.  The system is the
superfluid A-phase of liquid helium-3. While the right symmetry-breaking
pattern is present, it may be difficult to observe Cheshire charge phenomena
experimentally.  There are many complications in dealing with a real
condensed-matter system rather than a relativistic field theory.  Some of these
difficulties will be pointed out.

\subsection{The Order Parameter in Superfluid He-3 A}

He-3 atoms are fermions.  A condensation of Cooper pairs of atoms is thought to
be responsible for superfluidity in this system.  Unlike the electrons in BCS
superconductors, however,  the helium atoms tend to pair in p-wave, rather than
s-wave states,  so they have a net orbital angular momentum of 1.  In order for
the two-atom wavefunction to be symmetric, therefore, members of the pair must
also have their {\it spins} aligned in a triplet state with total spin 1.  In
addition to the overall phase of the condensate wavefunction, there are thus
two separate angular momentum vectors which can a priori rotate independently.
The full  internal symmetry group of the pair wavefunction is
\Ref\Salomaa{M.~M.~Salomaa, G.~E.~Volovik, Rev.~Mod.~Phys. {\bf 59} (1987),
533.}
\Ref\book{D. Vollhardt and P. W\"{o}lfle,  {\it The Superfluid Phases of
Helium-3\/} (Taylor and Francis, New York, 1990).}
$$SO(3)^{(L)}\times SO(3)^{(S)}\times U(1)^{\phi}. $$
This richness in degrees of freedom leads to a wealth of interesting phenomena
associated with He-3 superfluidity.
Let us denote the generators of these three factors by
 $\hat{\vec{L}},\hat{\vec{S}}$, and $\hat{I}$, respectively.
The symmetry of the superfluid ground state depends on temperature and
pressure:
there are at least two phases which are stable in bulk, unmagnetized fluid,
characterized by different ground state configurations.  The A phase is
described by a spin
state $|1 0>$ along some preferred direction,
and an orbital state $|1 1>$ along some other axis.
    The object corresponding to the Higgs field is a $3\times 3$ complex matrix
of two-particle correlation functions, $A_{ai}$, with each entry representing a
particular spin state and orbital harmonic.  Rotations in spin space act on the
first index, a, and rotations in ordinary space act on the second index:
$A_{ai}$
transforms as a vector under each of the two $SO(3)$ factors of the symmetry
group.  The $U(1)$ factor acts on the overall phase of the matrix.  In the
A-phase, the matrix takes a value of the form:
$$ A_{ai}=\Delta_{A}(T) d_{a}(e_{1i}+ie_{2i})e^{i\phi}\eqn\a$$
Here $\Delta_{A}(T)$, a temperature-dependent gap parameter, can be thought of
as the magnitude of the superfluid wave function, much like the magnitude of
the higgs vev in a field theory.
The vector $\d$ is the axis along which the projection of the
spin angular momentum is zero.  $\vec{e_{1}}$ and $\vec{e_{2}},$ together with
the vector
    $\l=\vec{e_{1}}\times \vec{e_{2}},$ define a local orthonormal frame such
that the projection of the pair's mutual {\it orbital} angular momentum onto
$\l$ is $+1$.
 The phase $\phi$ represents the overall phase of the pair wavefunction.

In order to see what the pattern of symmetry-breaking is, consider what
transformations leave the order parameter 
invariant.
Continuous rotations
in spin space about the axis $\d$ leave $A_{ai}$ unchanged. These form an
unbroken $U(1)$ subgroup  with generator $\hat{S}_{z}$ (The state $|1 0>$ is
invariant under rotations about the z axis.)
Rotations in orbital space about $\vec{l}$ result in a phase (corresponding
to the phase gained in rotations of the state $|1 1>$ about the z axis.)
However, this can be
compensated by a change in $\phi$.
Thus the $\hat{L}_{z}-\hat{I}$  generates another unbroken U(1) subgroup.
These are the only two elements of the Lie algebra which annihilate $A_{ai}$,
but the discrete transformation $\d\rightarrow -\d,\  \phi\rightarrow\phi+\pi$
leaves the matrix invariant.  The $\d$  vector can be flipped 180 degrees if
the phase $\phi$ is simultaneously shifted by $\pi$.  This makes the unbroken
group $H= U(1)\times U(1)\times_{S.D.} Z_{2}.$

Because of the presence of this discrete $180^\circ$
  rotation  in the little group, the spin quantization axis $\d$
acts like the director field in a NLC:  there are configurations in which $\d$
can be rotated continuously through $180^\circ$ along a closed path
which winds once around the core of a vortex. Such a configuration is the
``half-quantum vortex,'' so called because the phase $\phi$ winds only halfway
around the unit circle
and the vortex carries only half of the conventional quantum of circulation.
Half-quantum vortices are analogous to the Alice vortices of the previous
section.  A similar
zero mode should in principle exist.  In the next section,  configurations with
such a zero mode are described.

\subsection{Half-quantum Vortices in He-3 A}

Static configurations of the superfluid order parameter are extrema of the
Landau-Ginzburg free energy functional \refmark{\Salomaa}, which takes the
place of the field Hamiltonian.  The free energy density includes a gradient
energy term:
$$F_{G}=\gamma_{1}\partial_{i}A_{\alpha j}\partial_{i}A^{*}_{\alpha j}+
       \gamma_{2}\partial_{i}A_{\alpha i}\partial_{j}A^{*}_{\alpha j}+
   \gamma_{3}\partial_{i}A_{\alpha j}\partial_{j}A^{*}_{\alpha
i}\eqn\gradient$$
where $\gamma_{i}$ are constants.  For general values of $\gamma_{i}$, this
term is not invariant under rotations of the orbital frame
$(\vec{e_{1}},\vec{e_{2}},\l)$ unless the external coordinates are
simultaneously rotated.  However, since the spin indices $\alpha$ are never
contracted with any of the differentiation indices, the gradient energy is
invariant under all global rotations of $\d$, regardless of the values of
$\gamma_{i}$.  $SO(3)_{Spin}$ is truly  an internal symmetry if only the
gradient energy is included.

Only the spin-orbit, or dipole, interaction couples spin with orbital indices.
This term  has no analog in the model of section 2:
  $$ F_{D}=g_{D}(A_{ii}A^{*}_{kk}+A_{ik}A^{*}_{kl}) =
          -2g_{D}\Delta_{A}^{2}(\d\cdot \l)^{2}.\eqn\a$$
The dipole force is weak compared to the other interactions, but it has the
consequence that in the bulk fluid,  $\d$ tends to line up parallel to $\l$.
This is known as dipole locking.

In the presence of a pair of HQV's,  the dipole energy must depart from
minimum over some region between the cores (dipole unlocking).  This is because
$\l$, unlike $\d$,  cannot wind by odd multiples of $\pi,$ so $\l$ will tend
instead to remain constant.
When the cores are widely separated, the consequence is that the winding of the
$\d$ vector occurs within a domain wall, or soliton, whose width is of order
 $$\xi_{D}\sim\sqrt{\gamma_{i}\over{g_{D}}}.\eqn\a$$

$\xi_{D}$, known as the dipole length, is the scale at which the dipole energy
becomes comparable to the gradient energy.  The presence of a domain wall
causes the half-quantum vortices to be confined linearly rather than merely
logarithmically in two dimensions.  Figure
\FIG\DipoleSoliton{ Domain wall of dipole energy.  The spin axis $\d$ is
              represented
              by the undirected line segments, while the thick arrows represent
              $\l$.  the region in which $\l$ and $\d$ are not parallel has a
              width of order $\xi_D$.}\DipoleSoliton\/
shows $\l$ and $\d$ for such a
configuration.
Since the dipole energy depends only on the angle between $\l$ and $\d$  a
global rotation of all the d vectors in figure \DipoleSoliton\/
about the x axis will still leave the Landau-Ginzburg free energy invariant.
This is the zero mode which gives rise to Cheshire charge in this case: a
global rotation which belongs to the subgroup unbroken at infinity.

\subsection{The Observability of Cheshire Charge}

In the case of He-3 vortices, the ``charge'' that can be transferred is a form
of angular momentum.  The momentum conjugate to  $\vec{\rm d}$  is the spin
density, or net nuclear magnetization, $\vec{S}$.  In the A-phase equilibrium,
the spin density has  expectation value zero.  Dynamics slower than the gap
frequency characterizing the symmetry-breaking scale but faster than
$\sim 1 Hz$ is governed by an effective Hamiltonian
$$ {\cal H}=\half \gamma^{2}S_{\alpha}(\chi^{-1})_{\alpha \beta}S_{\beta}
            -\vec{H}\cdot \vec{S} +F_{G} + F_{D},  \eqn\Leggettham$$
where $\vec{H}$ is an externally applied magnetic field and
 $\gamma$ is the gyromagnetic ratio of the atomic spins.  Unless otherwise
stated, we will assume no external field.  $\chi$, the magnetic susceptibility,
 is a symmetric tensor with two distinct eigenvalues $\chi_{\parallel}$
 and $\chi_{\perp}$ reflecting the greater polarizability of the fluid in
directions perpendicular to d:
 $$\chi_{ij}=\chi_{\parallel}d_{i}d_{j}+
 \chi_{\perp}(\delta_{ij}-d_i d_j).\eqn\a$$
In spite of the coupling of $\l$ and $\d$ through the dipole interaction,
\Leggettham\/ ignores  the dynamics of the
orbital axis $\l$ is because the motion of $\l$ is so strongly damped that all
but very slow motions of $\d$ can be treated as occuring on a
background of fixed $\l$. \book
The Hamiltonian \Leggettham \/, together with the commutation relations
$\{S_i, d_j\} = \epsilon_{ijk}d_k,$\/ $\{S_i, S_j\} = \epsilon_{ijk}S_k,$
leads to the Leggett \Ref\Leggett{A.~J.~Leggett, Ann.~Phys. {\bf 85} (1974),
11.} equations of motion.  $\d$ precesses about $\vec{S}$ according to
$$ \partial \d /\partial t = \gamma \d \times (-{{\gamma
\vec{S}}\over{\chi_{\perp}}}). \eqn\precession$$
Thus the first term of \Leggettham \/ plays the role of a kinetic term for
motions of $\d$, with $\vec{S}$ being the
momentum.  In particular, for the pair configuration shown in Figure
\DipoleSoliton,  the $x$ component of $\vec{S}$ will be nonzero when the charge
rotor is excited, and the quantized zero mode will have excitations where the
total angular momentum $\int d^{d}x S_{x} = n \hbar$.
These excitations will carry an energy of order
$$ E_{ZM} \sim {\gamma^{2} {n^{2}\hbar^{2}}\over{\chi_{\perp} V_{\rm
Soliton}}},\eqn\a$$
where $V_{\rm Soliton}$ is the volume of the dipole-unlocked region near the
cores.
 For large $n$, one can identify a classical precession frequency
$\omega=(1/\hbar)dE/dn,$ and the kinetic energy is given by:
$$ E_{ZM} \sim {\omega^{2} V_{\rm Sol}\chi_{\perp}
\over{\gamma^2}}.\eqn\EZMomega $$

As in the model of Section 2, this energy has, for fixed $n$, an inverse
dependence on the soliton volume, and thus on the separation of the cores.  In
a two-dimensional geometry where string tension does not operate, it is
possible that a sufficiently excited pair could be stabilized against the
attractive force due to the dipole interaction, provided the ``Cheshire
charge'' cannot be radiated away rapidly.  This would occur if the energy
stored in the zero mode was comparable to the dipole energy of the the soliton
connecting the two cores:
$$ E_{D}\sim g_{D}\Delta_{A}^{2} V_{\rm Sol.}\sim  E_{ZM} \sim \gamma^{2}
{{n^{2}\hbar^{2}}\over{V_{\rm Sol}\chi_{\perp}}}.\eqn\CompareEnergy$$
Using formulas and numbers that can be found in Reference[\Salomaa], one
can estimate the volume at which this occurs (more details of this estimate
are found in the Appendix):
$$ V \sim n \times (1-T/T_c)^{-1/2}\times 10^{-16} cm^{3}. \eqn\estimate$$
The classical precession frequency of $\d$ corresponding to this energy level
is approximately $10^{4} s^{-1}$.   By comparison, the gap frequency
below $T_c$ is typically $\sim~kT_{C}/\hbar\sim 10^7~s^{-1}$.

The stability of a charged excited state of a vortex loop or pair is also
uncertain. It may depend on details of the spin relaxation behavior of the
fluid and such factors as coupling between the superfluid and the normal fluid
component\Ref\Cross{M.~Cross, private communication.}.
However,  since the precession frequency found above to be sufficient to cancel
the attractive force is lower than the gap frequency, one would imagine that at
least the radiation of ``Higgs'' modes would be suppressed. Also, since the
``Cheshire charge'' consists of a nonzero spin density in the direction of $\l$
(assuming that $\l$ maintains a uniform value everywhere which is parallel to
the asymptotic value of $\d$), no torque should be exerted on this
component of $\vec{S}$ by the dipole force.  This renders one of the usual
means \refmark{\book}  for relaxation of $\vec{S}$ ineffective: namely its
damping by coupling to $\l$.

\subsection{A Charge Exchange Process}

The spin wave excitations of the He-3 A order parameter carry quantum numbers
which allow the possibility of an Aharonov-Bohm interaction with half-quantum
vortices.  In general, spin waves consist of a coupled oscillation of $\d$ and
$\vec{S}$.  Consider the form
$$ \d = \d_{0}(\r) + \vec{\psi}(\r, t).$$
Oscillations of $\psi$ parallel to $\d_{0}$ have a gap characterized by the
symmetry-breaking scale. On the other hand, oscillations of $\psi$
perpendicular to $\d_{0}$ only have a frequency shift $\Omega_A$ proportional
to the dipole energy. In particular, the two propagating low-frequency modes
 obey a wave equation of the form
$$ -{{\partial^2 \psi} \over {\partial t^2}}=
   \Omega^{2}_{A}\psi + \Omega^{2}_{A}(U \psi + D\psi),\eqn\spinwaveq$$
where
$$\Omega^{2}_{A} = {\gamma^{2} 2 \Delta^{2}_{A} g_{D}\over
\chi_{\perp}},\eqn\a$$
$U$ is a potential which is zero in the bulk fluid, and D is a kinetic
operator:
$$D\psi=-\xi^{2}_{d}[\Delta\psi+
 {\rho^{\parallel}_{sp}-\rho^{\perp}_{sp}\over\rho^{\parallel}_{sp}}
 \del\cdot(\l(\l\cdot\del)\psi)].\eqn\kinetic$$

$\rho^{\perp}_{sp}\propto (2\gamma_1 + \gamma_2+\gamma_3)$
and $\rho^{\parallel}_{sp} \propto 2\gamma_1$ are the spin rigidity
coefficients
describing the energy of gradients in $\d$.
 The lower cutoff frequency $\Omega_{A}$ arises because an oscillation of $\d$
about $\l$ is an oscillation in the potential well formed by the dipole
coupling $g_D (\l\cdot\d)^{2}$.  (As mentioned previously, $\l$ can effectively
 be regarded as fixed on the time scales of these oscillations.  The
fluctuations of $\l$ are diffusive or overdamped.)  The potential U becomes
nonzero when $\l$
 is not parallel to $\d$  (as inside domain walls) or when nonuniform textures
 of the order parameter are present.
 These oscillations of $\psi$ perpendicular to $\d_{0}$ would be Goldstone
modes if the dipole energy were neglected, and they are
  degenerate with each other as long as $\d_{0}\parallel\l.$
\Ref\Wolfle{P~W\"{o}lfle, Physica {\bf 90B} (1977), 96.  G.E.~Volovik,
M.V~Khazan. Soviet Physics JETP {\bf 58} (1983), 551.}\/ There are thus three
modes which have the pattern of splitting analogous to the splitting  of the
$\Psi$ modes in section 2.

Figure \FIG\SpinDrag{Parallel transport of orthogonal spin wave modes.
               The effect of parallel transport about an HQV core on the two
               degenerate spin-wave modes is shown. $\d$ is indicated by the
               undirected line segments.  The amplitude of oscillation of the
               spin density $\vec{S}$ is shown by the thick arrows.  The thin
               arrows show the corresponding motion of $\d$.  In mode 1 (upper
               figure) the spin density amplitude points out of the page and
               remains the same on transport around the core.  For mode 2,
               (lower figure) the spin density amplitude is within the page
               and experiences a sign change. }\SpinDrag\/
 demonstrates the ``frame dragging''  of the spin wave modes.  We assume a
frequency lower than the gap frequency, which corresponds to the assumption in
Section 2 or in reference [\MPW] that scattering experiments are done at an
energy such that only the light components of the split multiplet $\Psi$ are
excited. Then the two propagating oscillations (corresponding to $\Psi_1$ and
$\Psi_2$)
are the two different polarizations of $\psi$ perpendicular to $\d_0$.  In
accordance with eqn.  \precession,  the fluctuation of $\d$ is accompanied
by a fluctuation of $\vec{S}$ in the $\vec{\psi}\times\d_0$ direction.  In the
region far from the vortex cores, $\l$ and $\d$ are taken to lie along the
x-axis, so that one of the propagating modes, labeled 1, involves $\psi_y$ and
$S_z$, while mode 2 involves $\psi_z$ and $S_y$.  As one follows a path around
one of the vortices, however,  mode 2 experiences a frame-dragging which causes
it to
mix with $S_z$, acquiring an Aharonov-Bohm minus sign when transported around a
loop.  Mode 1 remains unaffected.
It is therefore conceivable that a similar charge exchange process could occur
in the scattering of spin waves off pairs of half-quantum vortices.  A
circularly polarized spin wave (an eigenstate of the unbroken subgroup of
rotations) could scatter from the pair of vortices, changing to the opposite
circular polarization and depositing angular momentum (Cheshire charge) in the
vicinity of the vortices.

The theoretical possibility of an Aharonov-Bohm type scattering from HQV's
using collective excitations of the fluid as the projectiles has been mentioned
previously by Khazan and others \REF\Khazan{M.V.~Khazan. JETP Lett. {\bf 41}
(1985), 486.   M.~M.~Salomaa, G.~E.~Volovik. Phys. Rev. Lett. {\bf 56} (1986),
1313. } \refmark{\Khazan,\Salomaa}.
However, the context studied by these authors was that of NMR experiments in
which either spin waves or an orbital ``Higgs'' excitation called the clapping
mode
 are excited by means of a  fluctuating magnetic field.
The high {\it steady-state\/} magnetic field which is used in NMR breaks the
degeneracy between the two ``light'' spin wave modes, leaving us with an
abelian situation like that of reference [\MPW].  Situations in which the
non-abelian Aharonov-Bohm effect might be seen were not discussed.

It may in practice be difficult to devise a Cheshire charge experiment without
having the $SO(3)$ symmetry destroyed by an external field.  An additional
difficulty arises from the potentials U in equation \spinwaveq.  In addition to
the Aharonov-Bohm effect, one
expects spin waves to be scattered by these non-topological potentials which
are nonzero within a soliton where $\d_0$ is not parallel to $\l$.
  These potentials arise because of the change in the dipole restoring force as
$\d_0$ leaves the bottom of the dipole potential well,  and because of other
anisotropies associated with the orbital state.  In fact, the two linear
polarization
states of the spin wave experience different potentials inside the soliton,
which could give rise to phase shifts between them in addition to the
Aharonov-Bohm phase.

\Appendix{A}{Estimate of Charge Necessary to Stabilize Pair of HQV's.}

This appendix contains a derivation of the order-of-magnitude estimate
\oldeq{\estimate} \/ for the level of excitation of the charge rotor mode (and
corresponding
classical precession frequency) at which its energy becomes
 comparable to the dipole energy of the soliton connecting a pair of
half-quantum vortices. The data and formulas used here can be found in
references
 [\Salomaa] and [\book].

We begin with equation \oldeq{\CompareEnergy}, equating the dipole energy with
the zero-mode energy:
$$ E_{D}\sim g_{D}\Delta_{A}^{2} V_{\rm Sol.}\sim  E_{ZM} \sim \gamma^{2}
{{n^{2}\hbar^{2}\chi_{\perp}}\over{V_{\rm Sol}}}.\eqn\equality$$

$\chi_{\perp}$ is the larger eigenvalue of the magnetic susceptibility tensor.
It differs from the susceptibility $\chi_N^0 = \gamma^2 \hbar^2 N(0)$ of a
noninteracting degenerate Fermi gas only by a factor of order unity, so we may
use this value as an estimate of $\chi_{\perp}$.  In the previous expression,
$N(0)$  is the density of states at the Fermi surface, given by
$N(0)=m^{*}k_f/2\pi^2 \hbar^2$  where $k_f \hbar$ is the Fermi momentum.

We also rewrite the dipole coupling constant in terms of measurable length
scales as follows:
$$ g_D = {\gamma_0 \over \xi_D^2} = {N(0) \xi_0^2 \over \xi_D^2}.\eqn\a$$
$\gamma_0$ is a typical coefficient of the gradient energy:
in the weak-coupling, or small-interaction limit, the coefficients $\gamma_i$
in the gradient energy   \oldeq{\gradient} \/   are all equal to $\gamma_0$.
We have in turn related this coefficient to the coherence length
$\xi_0 (1-T/T_C)^{-1/2}$.  The two length scales are quoted by [\Salomaa] as of
order $\xi_D \sim 10^{-3} cm $ and $\xi_0 \sim 10^{-6} cm$.

Finally, we use the relation $\Delta_A \sim kT_{C} (1-T/T_C)^{1/2}$ for the gap
parameter.  We can substitute the estimates for $\chi_{\perp}$, $g_D$, and
$\Delta_A$ into \equality.  Assuming temperatures in the millikelvin
range, molar volumes of a few tens of $cm^3$, and quasiparticle mass
$m^{*}$ approximately equal to the atomic weight of helium,  we obtain:
$$  V_{\rm Sol} \sim {{n \hbar^2 \xi_D}\over{\xi_0 m^* k_f
kT_{C}(1-T/T_C)^{1/2}}}
     \sim n \times 10^{-16} cm^3. \eqn\a$$

We may also express the answer in terms of a classical precession frequency.
Expressed in terms of $\omega$,  \oldeq{\equality}\/ becomes:
$$ {{\omega^2 V_{Sol} \chi_{\perp}}\over{\gamma^2}} \sim g_D \Delta_A^2
V_{Sol}. \eqn\a$$
Using the same estimates as above, we find:
$$\omega \hbar \sim {\xi_0 \over \xi_D} \Delta_A.\eqn\a$$
This shows that the frequency is of order $10^{-3}$ times the gap frequency, or
about 10~kHz if $(1-T/T_C) \sim 1$.  Not surprisingly, this is also of the same
order as the cutoff frequency $\Omega_A$ for spin waves.

\ack
I would like to thank John Preskill, Robert Navin and Hoi-Kwong Lo for helpful
discussions.

\par\penalty-400\vskip\chapterskip\spacecheck\referenceminspace
   \ifreferenceopen \Closeout\referencewrite \referenceopenfalse \fi
   \line{\fourteenrm\hfil REFERENCES\hfil}\vskip\headskip
   \input referenc.txa
   
\par\penalty-400\vskip\chapterskip\spacecheck\referenceminspace
   \iffigureopen \Closeout\figurewrite \figureopenfalse \fi
   \line{\fourteenrm\hfil FIGURE CAPTIONS\hfil}\vskip\headskip
   \input figures.txa
   
\end